\documentstyle{mn}
\begin{document}


\newcommand{\beq}{\begin{equation}}
\newcommand{\eeq}{\end{equation}}

\def\msun{{\rm\,M_\odot}}

\def\bi{\bibitem}
%
\newcommand{\journ}[5]{
                      {#1},       
                      {#2}.       
                      {\it #3},   
                      {\bf #4},   
                      {#5}.       
                      }
%
\newcommand{\inpress}[4]{
                  {#1},        
                  {#2},        
                  {\it  #3\/}, 
                  {#4}.        
                        }
%
\newcommand{\inprep}[3]{
                  {#1},  
                  {#2}.  
                  {#3}.  
                       }
%
\newcommand{\book}[4]{
                   {#1},        
                   {#2}.        
                   {\it  #3\/}, 
                   {#4}.        
                   }
%
\newcommand{\proceed}[6]{
                   {#1},        
                   {#2}.        
                in {\it  #3\/}, 
              eds. {#4},        
                   {#5},        
                p. {#6}.        
                        }
%
\newcommand{\proceedinpress}[5]{
                   {#1},        
                   {#2},        
                in {\it  #3\/}, 
              eds. {#4},        
                   {#5}.        
                               }
%
\newcommand{\thesis}[3]{
                   {#1},        
                   {#2},        
                   {\it  #3\/}, 
                       }

%


\title[Final Stages of N-body Star Cluster  Encounters ]
{Final Stages of N-body Star Cluster  Encounters  }
\author[M.R. de Oliveira {\it et al.}]
{M.R.~de Oliveira,$^1$ E.~Bica,$^1$ H.~Dottori,$^1$\\
$^1$Instituto de Fisica-UFRGS, CP 15051, CEP 91501-970 POA - RS, Brazil}

\maketitle

\begin{abstract}

 We performed numerical simulations of star cluster encounters with
 Hernquist's TREECODE in a CRAY YMP-2E computer. We used different
 initial conditions (relative positions and velocities, cluster sizes,
 masses and concentration degrees) with total number of particles per
 simulation ranging from 4608 to 20480. Long term interaction stages
 (up to 1 Gyr) when the pair coalesces into a single cluster are
 compared with isolated LMC clusters. Evidence is found that when seen
 in a favourable plane these resulting clusters show elliptical shapes
 due to the disruption of one of the companions. These elliptical
 shapes are essentially time independent but they do depend on the
 initial structural parameters of the pair components.  We also
 analyzed the fraction of stars that are ejected to the field by the
 interaction.  We found that this fraction can be almost $50\%$ for
 the disrupted cluster. These simulations can represent a possible
 mechanism to explain the ellipticity observed on several star
 clusters in the Magellanic Clouds.

\end{abstract}


\section{Introduction}

The study of star cluster pairs in the Magellanic Clouds has been
explored extensively in the past years (Bhatia \& Hatzidimitriou 1988;
Bhatia \& McGillivray 1988; Bica, Clari\'a \& Dottori 1992; Rodrigues
{\it et al.}  1994, hereafter RRSDB; Bica \& Schmitt 1995; de Oliveira
{\it et al.} 1998, hereafter Paper I; Bica {\it et al.} 1999). The
evolution of physical pairs can provide fundamental insight into the
past history of cluster formation in the Magellanic Clouds.

In the last decade, N-body simulations of stellar system encounters
have been the main tool to investigate the dynamical processes that
can occur in such interactions, like mergers and tidal
disruption. Many of the first simulations worked with equal mass
encounters (White 1978; Lin \& Tremaine 1983; Barnes 1988). Different
mass encounters have been carried out by Rao, Ramamani \& Alladin
(1987), Barnes \& Hut (1986, 1989), RRSDB and Paper I. These studies
indicate that tidal disruption and merger are two important processes
in the dynamical evolution of a binary stellar system. However, there
remain important issues not yet well tested, as the long term stages
of such interactions, and the fraction and how stars are stripped from
the cluster by the parent galaxy tidal field.

Some isolated clusters in the Magellanic Clouds present interesting
structures with higher ellipticities than Galactic globular clusters
(van den Bergh \& Morbey 1984).  Several explanations have been
proposed like the presence of sub clumps merging with the main cluster
which could produce the impression of high ellipticities in these
clusters (Elson 1991). Another possibility is that old mergers in
advanced evolutionary stages could show such ellipticities (Sugimoto
\& Makino 1989).

This work is a continuation of Paper I where we analyzed the
morphology of selected cluster pairs in the Magellanic Clouds and have
compared them to those obtained from numerical simulations of star
cluster encounters. A preliminary discussion was given in de Oliveira
{\it et al. } (1999). In the present paper we study the long term
behaviour of some of our numerical models (up to 1 Gyr), and compare
the final stages of these simulations (isodensity maps, ellipticities,
isophotal twisting) with the structure of isolated clusters.

In Section 2 we describe the method and the conditions employed in the
present simulations. In Section 3 we describe the Magellanic Cloud
cluster images and the procedure to derive the isodensity maps and
ellipticity measures. The numerical results and discussions are
presented in Section 4. In Section 5 we give the main conclusions.

\section{The Method and the Initial Conditions}

We performed the simulations using TREECODE (Herquist 1987) in the
CRAY YMP-2E computer of Centro Nacional de Supercomputa\c c\~ao of the
Universidade Federal do Rio Grande do Sul (CESUP-UFRGS). For a
complete description of the method see Paper I.

\subsection{The Initial Conditions}

In Paper I, cluster N-body models were generated for 16384, 4096 and
512 equal mass stars, corresponding to total masses of $10^{5} \msun ,
10^{4} \msun$ and $10^{3} \msun$, respectively. The cutoff radii of
the Plummer clusters are 20, 15 and 8 parsecs, respectively. These
values are comparable to diameters found in Magellanic Clouds clusters
(Bhatia {\it et al.} 1991).  We generated two different concentration
degrees for the 512 particle cluster, one with half-mass radius 1/2 of
the cutoff radius and another with 1/4 (Table~\ref{tab:iniciais}). The
interaction models are characterized by a pericentre distance $p$
(distance of closest approach) and the initial relative velocity
$V_{i}$.  These initial velocities are comparable to the observed
random velocities in the LMC disk (Freeman, Illingworth \& Oemler
1983). The initial conditions (t=0) used in Paper I were: (i) the
large cluster (in the case of equal mass clusters, the more
concentrated one) is located at coordinates (X,Y)=(0,0); (ii) the
small cluster lies at a distance $r_{0} = 5\times r_{t}$ (where
$r_{t}$ is the tidal radius). Beyond this distance the tidal effects
are negligible. The initial relative velocity at this distance was
obtained from the two-body formulae. The position and velocities of
the particles during the encounters were computed in the
centre-of-mass frame of the entire system.  At various times, the
essential data containing the positions and velocities of all the
particles were stored for later analysis. The computation was stopped
when disruption of the cluster occurred or, in the case of open
orbits, when the relative separation of the clusters reached distance
$d>2r_{0}$. We point out that the softening parameter adopted for the interaction  is the smallest
value of the two cluster  models  (column 7 of Table 1). Indeed, with reference to the  AB cluster
 interaction (which has the largest $\epsilon$ difference) we  carried out a simulation where cluster B is allowed to evolve isolated adopting the cluster A softening
 value for $\Delta t\sim$200 Myr ($\sim$ 2 relaxion times). Analysis of the time evolution  of
 cluster B structure shows that it is essentially the same for $\epsilon$ = 0.30   as that obtained for the cluster with  $\epsilon$ = 0.47. Consequently in the AB cluster interaction significant structural
changes are not expected  by the adoption of the smaller $\epsilon$ value. 
 
In the present  Paper, we proceeded the simulation from this point on
 for some models with closed orbit where disruption occurred (see
 Table~\ref{tab:colisao}). We used the same procedures as in Paper
 I. These simulations were  ran up to 990 Myr for the models presented here.

 \begin{table}

\caption{Input parameters for the simulations, by columns: (1)  number of particles of the cluster; (2)  cluster 
total mass; (3)  half-mass radius; (4)  maximum radius of the cluster; (5)
 mean velocity of the particles (modulus) and (5)   softening parameter. }
\label{tab:iniciais}
\begin{center}
\begin{tabular}{l r r r r r r}
\hline
\hline
Model & $N_{part}$ & \multicolumn{1}{c}{$M_T$} & \multicolumn{1}{c}{$R_h$} & \multicolumn{1}{c}{$R_{max}$} & \multicolumn{1}{c}{$V_{md}$} & $\epsilon$ \cr
      &            &   $(\msun)$   &   (pc)          &   (pc)        &      (Km/s)         &          (pc)                 \cr
\hline
A     & 16384    & $10^5$           &  4.94     &  20.0     &  3.53       &  0.30 \cr
B    &  4096    & $10^4$           &  3.72     &  15.0     &  1.94        &  0.47 \cr
C    &   512    & $10^3$           &  2.00     &   8.0     &  0.83        &  0.50  \cr
D    &   512    & $10^3$           &  4.00     &   8.0     &  0.62        &  0.50 \cr
\hline
\hline
\end{tabular}
\end{center}
\end{table}

We illustrate in Fig.~\ref{fig:simul_xy} the time evolution of an
 elliptic orbit encounter involving two clusters with 16384 and 4096
 particles respectively (model {\bf E9AB10}, see
 Table~\ref{tab:colisao}). In this Fig. we can see the small cluster
 being deformed by the massive one with the formation of a bridge ($\simeq$50
 Myr), subsequently complete disruption occurs ($\simeq$135 Myr); at the end
 of this simulation the clusters coalesce into a single one, with the
 stars of the disrupted cluster forming a halo around the final
 cluster, and some of them being ejected.

\begin{figure*}
\vspace{18cm}
\caption{Time evolution of an encounter (model E9BD10) projected in the XY plane. }
\label{fig:simul_xy}
\end{figure*}

\begin{table}
\caption{Collision parameters. By columns: (1) elliptic orbit model; (2)
 total number of particles of the two clusters; (3) mass ratio; (4)  eccentricity of the orbit; (5)   pericentre of the orbit; (6)  initial
relative velocity .}
\label{tab:colisao}
\begin{center}
\begin{tabular}{ l r r r r r r }
\hline
\hline
Model & $N_{part}$ & $M_{1}/M_{2}$ & $r_0$(pc) & {\it e} & {\it p}(pc) & $V_i$
(Km/s)\cr
\hline              
E9AB10    &  20480   & 10            &  70.0   &  0.9    &  20.0  & 4   \cr
E9BC10  &   4608   & 10            &  37.0   &  0.9    &  10.0 & 2       \cr
E9BD10    &   4608   & 10            &  37.0   &  0.9    &  10.0  & 1     \cr
E7BD10   &  4608   &  10           &   37.0  &  0.7    &  10.0  & 1       \cr
E6BD10  &   4608   &  10           &  37.0   &  0.6    &  10.0   & 1   \cr   
\hline
\hline
\end{tabular}
\end{center}
\end{table}

\section{Isodensities for LMC Elliptical Clusters }

In the final stages of our simulations the pairs coalesced into a
single cluster with a distinct structure as compared to the original
ones. We noticed that in  viewing angles close to the  original
orbital plane of encounter, the final single cluster presented an
ellipticity larger than those of the initial clusters (our models have
a spherical symmetry).  In this paper we have selected some LMC
clusters with morphologies resembling those of the present models
(Sect. 4.4). We checked many clusters with important ellipticity, as
indicated in previous studies (Geisler \& Hodge 1980, Zepka \& Dottori
1987, Kontizas et al. 1989, 1990).  In particular the selected
clusters present increasing ellipticity outwards. In
Table~\ref{tab:age} we show these clusters with their V magnitudes,
SWB types derived from interpreted UBV photometry (Bica {\it et al.}
1996, Girardi \& Bertelli 1998) and corresponding ages. Also there are 
available ages from colour-magnitude diagrams: (i) NGC1783 has 700-
1100 Myr depending on adopted distance modulus (Mould {\it et al.}
1989). Based on the same data, using the $\Delta$V turnoff/clump
method, Geisler {\it et al.} (1997) obtained 1300 Myr; (ii) NGC1831
has an age of 400 Myr according to Hodge (1984). CCD data provided
500-700 Myr (Mateo 1988) and 350-550 Myr (Vallenari {\it et al.}
1992), depending on adopted models; (iii) NGC2156 has age $\sim$60 Myr
according to Hodge (1983); (iv) NGC1978 has an age $\simeq$2000 Myr
(Olszwewski 1984, Bomans {\it et al.} 1995). Geisler {\it et al.}
(1997) derived 2000 Myr by means of the $\Delta V$ turnoff/clump
method.  There is good agreement of ages derived from the CMD studies and the integrated colours (Table~\ref{tab:age}). NGC1783 and NGC1831 have ages comparable to the long term
stages of our models, while NGC2156 falls somewhat short, and NGC1978 has an age $\simeq$ a factor 2 larger.

\begin{table}
\caption{Ages for the clusters in the present sample. By columns: (1) cluster name; (2) observed visual magnitude; (3) SWB type; (4)  age.}
\label{tab:age}
\begin{center}
\begin{tabular}{ l r r r  }
\hline
\hline
Name & \multicolumn{1}{c}{ V} & \multicolumn{1}{c}{ SWB} & \multicolumn{1}{c}{Age}  \cr
      &    &   type    &               (Myr)        \cr
\hline
NGC1783  &   10.93  &   V       &   900    \cr
NGC1831  &   11.18  &   IVA       &   400      \cr
NGC1978 &   10.70  &    VI     &  3000   \cr
NGC2156 &   11.38  &    III     &  140   \cr 
\hline
\end{tabular}
\end{center}
\end{table}

The images of this selection were obtained from the Digitized Sky
Survey (DSS). The plates are from the SERC Southern Sky Survey and
include IIIa-J long (3600s), V band medium (1200s) and V band short
(300s) exposures. The PDS  pixel values correspond to photographic
density measures from the original plates, and are not calibrated.

The digitized images, similary to those generated by the models, were
treated with the IRAF\footnote{IRAF is distributed 
by the National Optical Astronomy Observatories, which is operated by the
Association of Research in Astronomy, Inc., under cooperative agreement with the National Science Foundation, U.S.A.}  package at the Instituto de F\'{\i}sica --- UFRGS, applying a 2-d Gaussian filter to smoothen out individual stars, and to create isodensity maps. The ellipticity measurement was made with
the IRAF package task Ellipse which fits elliptical isophotes to data
images with the use of a Fourier expansion:

\beq
y  = \sum_{n=0}^{4} {A{_n}sin(n E)  +  B_{n}cos(n E)}
\eeq

where E is the position angle.  The amplitudes ($A_{3}, B_{3}, A_{4},
B_{4}$), divided by the semi-major axis length and local intensity
gradient, measure the isophotal deviations from perfect ellipse. Note that the IRAF routine requires 
that one indicates a first guess for the object major axis position angle. Subsequently the routine iteractively fits the best solution for each isophote.  We
stored the coefficient $B_{4}$ which tells whether the ellipse is disk
(a positive $B_{4}$ parameter) or box-shaped (a negative $B_{4}$
parameter), Bender {\it et al.} 1989. We also stored the semi-major
axis position angle variation. It should be noted that in general
ellipticity does not present simple behaviour.  Ellipticity varies
with radius (Zepka \& Dottori 1987, Kontizas et el. 1989,1990) and so
it is difficult to assign one to a cluster. We measured elliptical
isophotes starting at $R_h$ and stopping at the last possible non
diverging ellipse. In the models, this last ellipse occurred around
the radius containing 90\% of the cluster mass.

\section{Discussion}

We considered models in which the less massive cluster (perturbed) is
allowed to move in elliptic orbits around the more massive one
(perturber). The perturbed cluster is assumed to move in an orbit of
eccentricity $e=0.6,0.7,0.9$.  The values of the pericentre and the
eccentricity are meaningful only if the clusters are assumed to be
point masses moving in Keplerian orbits. However, soft potential
orbits are not conical and our use of these definitions is meaningful
only to a good approximation, but not strictly (White 1978).

The collision parameters of the simulations are given in
Table~\ref{tab:colisao}. The model designation (column 1 of
Table~\ref{tab:colisao}) contains information on the encounter
conditions. E refers to an elliptic orbit encounter; the number
following the letter E is related to the orbital eccentricity (column
5 of Table~\ref{tab:colisao}); A, B, C and D refer to the model type
(Table~\ref{tab:iniciais}); the last number is the orbit pericentre
(column 6 of Table~\ref{tab:colisao}).

From the analysis of the time evolution of each model as illustrated
 in Fig.~\ref{fig:simul_xy} for {\bf E9AB10}, it is possible to
 observe the trend to form bridges in the beginning of the encounters
 (see also Paper I). As time goes by the smaller cluster is completely
 disrupted by the massive one, resulting in a single cluster at the
 end of the simulation.

\subsection{The Structure of the Final Stages}
   
In Fig.~\ref{fig:simul_xz}, we illustrate the same encounter as in
Fig.~\ref{fig:simul_xy}, but in a different plane projection
(parallel to the orbital plane). We can observe the disruption of
the smaller cluster, with its stars occupying orbits around the
massive cluster. As these stars remain preferentially orbiting in the
same plane of the original encounter, they introduce an elliptical
shape to the final single cluster, when seen in a favourable plane
projection.

\begin{figure*}
\vspace{18cm}
\caption{Time evolution of an encounter (model E9BD10) projected in the XZ plane. }
\label{fig:simul_xz}
\end{figure*}

In Fig.~\ref{fig:simul_E9AB10} we show isopleths, ellipticity,
position angle and $4^{th}$ cosine coefficient (hereafter $B_{4}$ )
for the model {\bf E9AB10} in a XZ projection. These measures were
made for four different evolution times, the first panel being the
initial massive cluster {\bf A} before interacting (t=0 Myr). We can
observe a radial variation of ellipticity during the interaction,
which has also been observed for all other models. It also can be seen
a general trend of decreasing ellipticity toward the inner parts of
the models.  Zepka \& Dottori (1987) and Kontizas {\it et al.} (1989)
have observed internal variations of ellipticity in a number of LMC
clusters, finding a general tendency of increasing ellipticity towards
the inner parts of these clusters. However, they reported a few
exceptions ($\simeq 5\%$ of the data) which in turn have counterpart
in our models, and could be the result of an interaction.  Indeed
there are about 300 cluster pairs among a total population of 7847
extended objects in the SMC, inter-Cloud region and LMC (Bica \&
Schmitt 1995 and Bica {\it et al.} 1998). At least 50\% of the pairs
are expected to be interacting (Bhatia \& Hatzidimitriou 1988) and we
need a favourable projection (nearly edge-on orbits) to observe
elliptical structure like those of the models.

\begin{figure*}
\vspace{17cm}
\caption{In each panel we have the isopleth, the  ellipticity,  position angle and
$B_{4}$   coef.  for the model  {\bf E9AB10}, in a XZ projection. These measures were made
 for four different evolution time as indicated. The first stage shows the initial massive
 cluster {\bf A} before interacting.}
\label{fig:simul_E9AB10}
\end{figure*}

\begin{figure*}
\vspace{10cm}
\caption{Same as Fig.~\ref{fig:simul_E9AB10} for other two
models at the  final stage (990 Myr). }
\label{fig:simul_E9BD10}
\end{figure*}

In Fig.~\ref{fig:simul_E9AB10} there is little time variation of
ellipticity during the lifetime of our model. This variation occurs
mainly in the outer parts of the cluster, probably by the continuous
escaping stars of the disrupted cluster.

There occur small position angle variations
(Fig.~\ref{fig:simul_E9AB10}), which indicate that the elliptical
shape of the cluster stands out in the plane of the interacting
encounter. The $B_{4}$ coefficient clearly has a positive peak, which
is a sign of the presence of a disc component structure in the
resulting model. This suggests that the elliptical shape in our final
clusters is due to a rotation disk formed mainly by stars of the
disrupted cluster.

In order to check whether mass or size could affect the behaviour of
our models, we did the same analyses for models with different mass
and concentration degree, but with same orbital encounter. In
Fig.~\ref{fig:simul_E9BD10} we present the same measures of
Fig.~\ref{fig:simul_E9AB10} for other two models at the final stage
and we see similar ellipticity variations in radius. In order to
compare absolute ellipticity variations between models, and also
between different time stages in the same model, it is required to
define ellipticity at a common radial distance. As we have the main
shape changes due to variations in the outermost parts of these
clusters we decided to measure the ellipticity as the mean between
3$R_h$ and the outermost ellipse measured.

In Table~\ref{tab:elipses} we give the mean ellipticity between 3$R_h$
and the outermost ellipse $e_{out}$ for all the models in different
time stages. We conclude that in  model {\bf E9AB10} $e_{out}$ shows a small tendency of increasing with time; differences among the models are found
mainly in the shape of {\bf e} versus radial distance curve.

When comparing ellipticity between different models at the same appropriate time
 (Table~\ref{tab:elipses}), we observe a small reduction of $e_{out}$ from a
 more massive model ({\bf E9AB10}) to a less massive model ({\bf E9BC10}). This
 relation (ellipticity) versus (mass) has been suggested for the LMC clusters
 (van den Bergh \& Morbey 1984) where high mass clusters have higher angular
 momentum or have more difficulty in shedding it than do low mass
 clusters. However, when comparing a more massive model {\bf (E9AB10}) with a
 less massive one ({\bf E9BD10}) --- where we have in the last a different
 concentration degree for the disrupted cluster --- we observe no significant
 variation in $e_{out}$. Thus, the initial concentration degree affects the
 final shape of the resulting cluster. This suggests that a more concentraded
 cluster has a deeper potential well, making more difficult stripping its stars
 by the more massive cluster. On the other hand, stripping stars in the less
 concentrated cluster is more efficient, resulting in a more pronounced halo
 expansion. This is supported by the results in Table 5. This halo expansion
 when seen in an edge-on view seems to be an important factor to stablish the
 shape of the final merger.

We also compared the possible ellipticity variation between models
with equal mass and concentration degree, but different initial
orbital parameters. We observed no significant differences in
$e_{out}$ between the models.

\begin{table*}
\caption{Mean ellipticity measured between 3 $R_h$ and the outermost
ellipse for the models in different time scales. By columns: (1)
model; (2) to (9) ellipticity for different evolutionary times and
respective mean errors.}
\label{tab:elipses}
\begin{center}
\begin{tabular}{l r r r r r r r r r }
\hline
\hline
Model  & time  &   e  & $e_{error}$ & time & e & $e_{error}$ & time & e & $e_{error}$ \cr
       & (Myr)  &       &         & (Myr) &     &         &  (Myr) &    &     \cr
\hline
E9AB10 &  225  &   0.12  &  0.008  & 500  &  0.13 & 0.005  & 990  &  0.15 &  0.008 \cr
E9BC10 &   -   &   -     &   -     & 500  &  0.10 &  0.01  & 990  &  0.08 & 0.009   \cr
E9BD10 &   -   &   -     &    -    & 500  &  0.14 &  0.01  & 990  &  0.14 &  0.004  \cr 
E7BD10 &   -   &     -   &     -   & 500  &  0.14 &  0.01  & 990  &  0.13 & 0.009 \cr 
E6BD10 &   -   &     -   &    -    & 500  &  0.16 &  0.006 & 990  &  0.14 & 0.004 \cr  
\hline
\hline
\end{tabular}
\end{center}
\end{table*}

\subsection{Rotational Velocity Field }

In Fig~\ref{fig:vel} we show the line-of-sight rotational velocity field
($V_{y}$) for the model E9AB10 along the major axis at 990 Myr seen in a XZ
projection. The merger consists approximately of a rigid rotation core and an
outer halo ( $r\geq 20$ pc) with a Keplerian fall. The rotational velocity along
the major axis has a peak value of about 1 $Km s^{-1}$ at a radius of $20-30
pc$. It is interesting to see that the rigid rotation extends exactly to the
radius of the initial massive cluster before the encounter. This indicates that
the main contribution to the velocity field for $r > 20$ pc is due to stars of
the small cluster. This becomes evident when we plot the two star cluster  
member sets for a merged final model (Fig.~\ref{fig:plot_xz}) in a XZ projection. In this Fig. we plot the model EAB10 at T=990 Myr  and a reference box (side of 40 pc) centred at the more massive cluster. It is clear that the main contribution to the final composite system for $r > 20$ pc is due to stars of the small cluster.
Velocity distributions obtained with high
resolution spectra with large telescopes might reveal such rotation
curves, while in turn could be a signature of evolved stages of
cluster merging.

\begin{figure}
\vspace{8cm}
\caption{The line-of-sight rotational velocity field along the major
axis of the model E9AB10 at 990 Myr. The cluster is seen in a
``edge-on'' projection (XZ plane) where the abscissa is the X axis. }
\label{fig:vel}
\end{figure}

\begin{figure}
\vspace{8cm}
\caption{For the merger EAB10 at T=990 Myr we separete the stars belonging to
each cluster. The left panel shows the stars of the more massive cluster, and
the right one stars of the less massive in a XZ plane projection. We also plot a 40 pc side box centred at the more massive cluster as a reference.  }
\label{fig:plot_xz}
\end{figure}

\subsection{ Mass Loss}

In Table~\ref{tab:raio} we give the radii containing 10\%, 50\% and
90\% of the mass for the massive cluster in each interaction (model
{\bf A} or {\bf B}) for $t=0$ and for the final cluster stage. We can
observe that $R_{50}$ varies little for the initial and final cluster
stage when comparing models with same initial mass, suggesting that
the massive one has its original stellar content little affected by
the collision. However model {\bf E9AB10} shows a contraction of the
core, showing that the initial mass may affects the cluster final
structure. The radius containing 90\% of the mass is larger for all
our models, which shows that the final cluster outer halo is formed
mainly by stars from the disrupted cluster. The latter result is in
agreement with those presented in Paper I for the radial mass
distribution. There we showed that the maximum expansion for the
disrupted cluster in {\bf E} models always occurs in the plane of the
encounter and is a minimum in the {\bf z} direction. When we compare
the projected density distribution between the models before
interacting and at the end of the simulation we see that they are
similar within the region $r<10$ pc (Fig~\ref
{fig:perfil_dens}). However, the final cluster extends with a
dependence of $\rho\sim r^{-3}$.

\begin{table}
\caption {Radii containing 10\%, 50\% and 90\% of the mass before and
at the final stage of the encounters (in parsecs).}
\label{tab:raio}
\begin{center}
\begin{tabular}{l r r r }
\hline
\hline
Model & $R_{10}$ & $ R_{50}$  & $ R_{90}$  \cr
\hline
A (t=0)     & 2.03    & 4.94      &  11.79      \cr
B (t=0)   &  1.55    &   3.71    & 8.89      \cr
E9AB10 (t=990 Myr)   &  1.00       &   3.44        &  17.62       \cr
E9BC10 (t=990 Myr)   &  1.34       &  3.89         &  12.83         \cr
E9BD10 (t=990 Myr)   &   1.33      &  3.79       &    13.77     \cr
E7BD10 (t=990 Myr)   & 1.35     &    3.86    &    13.94     \cr
E6BD10 (t=990 Myr)   &  1.29   &  4.01     &    14.13     \cr
\hline
\hline
\end{tabular}
\end{center}
\end{table}

Together with changes in the internal structure of our final single
cluster, the present simulations suggests that a fraction of cluster
stars may be ejected to the field due to the encounter. In order to
quantify this fraction, we calculated the total energy per particle,
identifying those that had positive total energy which are expected to
escape the system.

In Table~\ref{tab:var_mass} we have for each model the final fraction
of stars that escape the cluster (in units of solar mass). The mass
fraction that is ejected to the field is very small, considering the
ensemble of the pair ($\leq$ 3\% ), so the contribution to the field
stars by a cluster pair encounter is not very significant. It should
be noted that when we study the merger of the cluster pair we are not
taking into account the tidal field of the parent galaxy. This should
truncate the outer halo of the final cluster, which would contribute
with an increasing fraction of stars ejected to the field. Adopting a
tidal radius $r_{tidal}\sim 65 $pc (distance moduli of 18.5 for LMC,
Westerlund 1990) for observed cluster's halo (Elson {\it et al.},
1987), we can estimate how much mass is beyond this limit. For the
model E9AB10, we estimate $M\sim 50\%$ of the mass of the disrupted
cluster (4.5\% of the total mass) is beyond this limit. So, it can be
concluded that although the total mass loss for the pair is not very
significant, the assumption of a maximum radius for the final cluster,
implies that half of the disrupted cluster stars feed the field.

\begin{figure}
\vspace{8cm}
\caption{Projected density distributions for the model E9AB10 at 990 Myr  (dashed curve) and for the model A before the interaction (solid curve). The radius is in parsecs and the density is in $M_{\odot}.pc^{-2}$.  }
\label{fig:perfil_dens}
\end{figure}

\begin{table}
\caption{Final cluster escaping mass fraction. By columns: (1) model;
 (2) mass fraction (3) absolute mass loss.}
\label{tab:var_mass}
\begin{center}
\begin{tabular}{l r r }
\hline
\hline
Model  &    mass fraction   &  absolute mass   \cr
       &        (\%)            &   ($\msun$)         \cr
\hline
E9AB10 &    2.90    &  3195    \cr
E9BC10 &    2.88      &   316  \cr
E9BD10 &    3.20       &   352  \cr 
E7BD10 &    1.85      &    204  \cr 
E6BD10 &    1.15      &    127    \cr  
\hline
\hline
\end{tabular}
\end{center}
\end{table}

\subsection{Comparison of Simulations with Magellanic Cloud Clusters}

Isopleth maps of projected planes at a given time $t$ of a suitable
model can be compared to the observed isodensity maps of selected
Magellanic Cloud clusters to infer on their dynamics.  In Paper I we
have used this method to search for evidences of interacting pairs. In
the present study, we use model isopleths together with other measured
parameters in order to compare our simulation final stages with
possible observational evolved products of cluster pair interaction.
Some examples are given in
Figs.~\ref{fig:comp1},~\ref{fig:comp2},~\ref{fig:comp3}
and~\ref{fig:comp4}, where we show some isolated LMC clusters
isophotes, and radial dependence of ellipticity, position angle and
$B_{4}$ coefficient.
 
The examples shown here have an ellipticity radial variation with a
trend to increase towards the outer parts. NGC 1783 has an ellipticity
curve compatible with our model {\bf E9BD10} when seen in a XZ
projection at 990 Myr (Fig.~\ref{fig:simul_E9BD10}, right). The cluster age (Sect. 3) is fully compatible
with the model evolutionary time. This scenario can be explained by
the result of a cluster pair encounter with the subsequent merger of
the pair in the early history of NGC1783.

NGC1831 has many resemblances with the models. The ellipticity
(Fig.~\ref{fig:comp2}) increases outwards, but in the external
isophote the value decreases again. This is observed in the model {\bf
E9BC10} at 990 Myr (Fig.~\ref{fig:simul_E9BD10}, left). The cluster presents
a considerable isophotal twisting like in the model.

As pointed out in Sect. 3, the age of NGC2156 falls short of the long
term stages of the models. But notice that asymmetrical early stages
(Fig.~\ref{fig:simul_xz}) at $\simeq$ 25  Myr , if seen favourably in
a given direction might present ellipticity variations. The negative
B4 coefficient behaviour suggests that if the cluster is indeed a merger,
disruption has not yet occurred to create a disc shape.

NGC1978 is an interesting intermediate age cluster due to the
pronounced ellipticity (Fig.~\ref{fig:comp4}), which is already
presented in the innermost available isophote. No isophotal twisting
occurs in this case. Recently Kravtsov (1999) searched for special
variations in the colour-magnitude diagram of NGC 1978. They found
evidence of some variations which they attributed to a possible
metallicity spread. Bomans {\it et al.} (1995) did not find any
evidence of age variation within the cluster. A merger scenario does
not require differences in age and/or metallicity. Indeed two clusters
born in the same association are expected to have a close orbit
encounter, which is the more favourable case for interactions (Paper
I).

The $B_{4}$ coefficient has different behaviours in the 4 clusters,
and in some cases has varying values in a given cluster. We might be
witnessing clusters with either discoidal (positive values) or boxy
(negative values) shapes. The average value of $B_{4}$ for NGC1783
suggests an overall boxy shape. NGC2156 is also boxy, but with a
systematic trend to become disc-shaped in the outer parts. NGC1831
shows a predominant positive $B_{4}$ suggesting a disc-shape. Finally,
NGC1978 on the average is positive thus indicating a predominantly
disc-shape. However the cluster has two distinct $B_{4}$ value
regions, 20$\leq r \leq$50 is definitively disc while r $>$ 50 should be
classified as boxy. Behaviours like this  have counterparts in
elliptical galaxies (see e.g. the sample of Goudfrooij {\it et al.}
1994). The $B_{4}$ coefficient is a promising tool to explore the
possibility of merger in galaxies or star clusters.

\begin{figure}
\vspace{8.5cm}
\caption{In each panel we have the isophote, the ellipticity, the
position angle and the $B_{4}$ coef.  for the cluster NGC 1783.}
\label{fig:comp1}
\end{figure}

\begin{figure}
\vspace{8.5cm}
\caption{In each panel we have the isophote, the ellipticity, the
position angle and the $B_{4}$ coef.  for the cluster NGC 1831.}
\label{fig:comp2}
\end{figure}

\begin{figure}
\vspace{8.5cm}
\caption{In each panel we have the isophote, the ellipticity, the
position angle and the $B_{4}$ coef.  for the cluster NGC 2156.}
\label{fig:comp3}
\end{figure}

\begin{figure}
\vspace{8.5cm}
\caption{In each panel we have the isophote, the ellipticity, the
position angle and the $B_{4}$ coef.  for the cluster NGC 1978.}
\label{fig:comp4}
\end{figure}

\section{Conclusions}

We used N-body simulations to study the final morphology and structure
 of star cluster pair encounters. We also compared these morphologies
 with those of elliptical isolated LMC clusters. The main conclusions
 may be summarized as follows:

\begin{enumerate}

\item Close orbit encounters of cluster pairs can lead to a single
 final cluster with a distinct structure of the original ones. When
 seen in a favourable plane projection, models show an elliptical
 shape comparable to those of some observed isolated LMC clusters.  This
 suggests that tidal encounters could be a mechanism to explain the
 ellipticity of several clusters in the Magellanic Clouds.

\item Evolved stages appear to be stable after $\simeq$ 200 Myr,
suggesting that the resulting ellipticity is not transient. The
simulations indicate that the initial concentration degree and mass
affect the final shape of the resulting cluster. In the final stages
all models present halo expansion, and some present core contraction.

\item The models show   mass loss for the composite
system.The fraction of stars ejected to the field by the encounter, is
not so significant ($\sim 3\%$ with respect to the sum of both
clusters). However it can represent up to $\sim 50\%$ of the stars of
the disrupted cluster, if we assume a tidal radius of $r_{tidal}\sim
65$ pc for the final merged cluster.

\item The velocity distribution of some final stage models presents
characteristic velocity patterns in favourable plane projections,
which could be used as an  observational constraint to test the present
scenarios.

Finally, we call attention that the merging of spheroidal systems can produce disc-shaped products. Thus not only boxy-shaped systems might be related to mergers.

\end{enumerate}

{\bf ACKNOWLEDGMENTS}

\medskip

 We thank Dr. Hernquist for allowing us to use TREECODE and the
 CESUP-UFRGS for alloted time in the CRAY YMP-2E computer. We are also 
 grateful to an anonymous referee for interesting remarks. We
 acknowledge support from the Brazilian Institutions CNPq, CAPES and
 FINEP.  The images in this study are based on photographic data
 obtained using the UK Schmidt Telescope, which was operated by the
 Royal Observatory Edinburgh, with funding from the UK Science and
 Engineering Research Council, until 1988 June, and thereafter by the
 Anglo-Australian Observatory.  Original plate material is copyright
 by the Royal Observatory Edinburgh and the Anglo-Australian
 Observatory. The plates were processed into the present compressed
 digital form with their permission.  The Digitized Sky Survey was
 produced at the Space Telescope Science Institute under US Government
 grant NAG W-2166

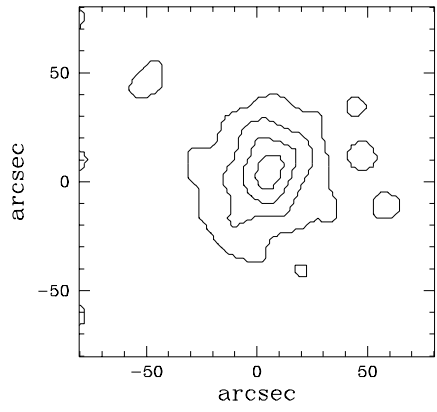
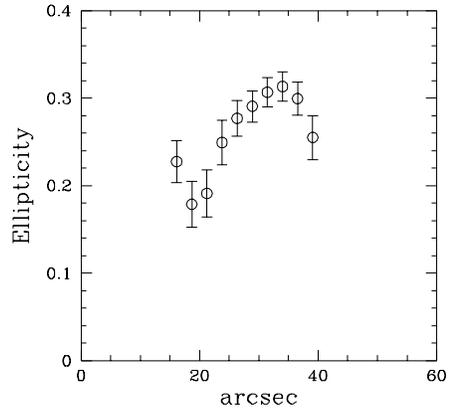
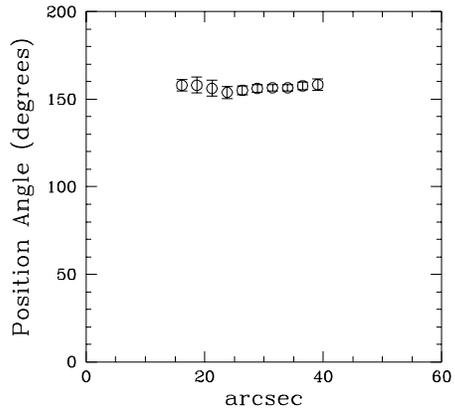
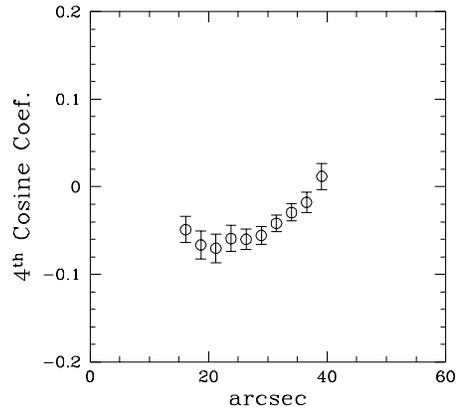
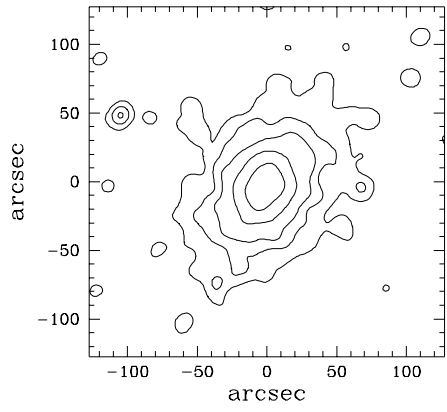
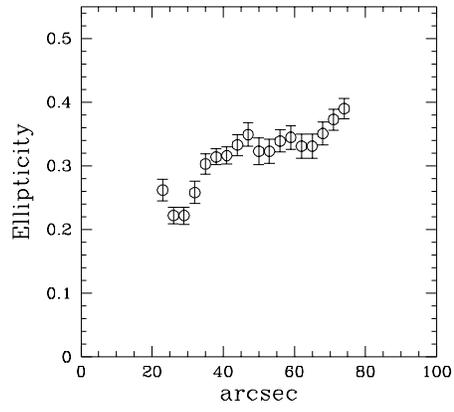
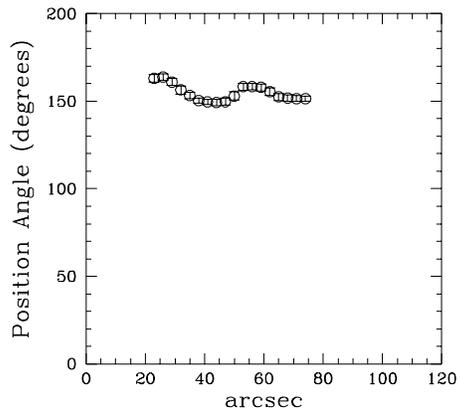
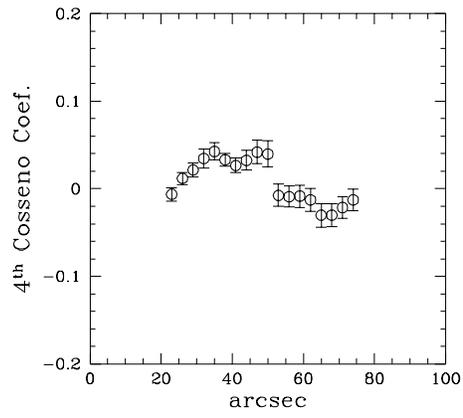
\end{document}